\documentclass[aps,prd,superscriptaddress,floatfix,nofootinbib,preprintnumbers]{revtex4}
\usepackage{amsmath,multirow}
\usepackage{graphicx,tabularx,slashed}
\usepackage{url}
\usepackage{footmisc}

\newcommand{\sg}{\ensuremath{G}}

\newcommand\one{\leavevmode\hbox{\small1\normalsize\kern-.33em1}}

\newcommand{\lag}{\mathcal{L}}

\newcommand{\qqquad}{\qquad \qquad}
\newcommand{\qqqquad}{\qquad \qquad \qquad}

\newcommand{\msbar}{\ensuremath{\overline{\text{MS}}}}

\newcommand{\gev}{{\ensuremath\rm GeV}}
\newcommand{\tev}{{\ensuremath\rm TeV}}

\newcommand{\pb}{{\ensuremath\rm pb}}
\def\slashchar#1{\setbox0=\hbox{$#1$}           % set a box for #1
   \dimen0=\wd0                                 % and get its size
   \setbox1=\hbox{/} \dimen1=\wd1               % get size of /
   \ifdim\dimen0>\dimen1                        % #1 is bigger
      \rlap{\hbox to \dimen0{\hfil/\hfil}}      % so center / in box
      #1                                        % and print #1
   \else                                        % / is bigger
      \rlap{\hbox to \dimen1{\hfil$#1$\hfil}}   % so center #1
      /                                         % and print /
   \fi}

\def\eg{{\sl e.g.} \,}
\def\ie{{\sl i.e.} \,}

\newcommand{\myrbox}[1]{\parbox{2.5cm}{\resizebox{2.5cm}{!}{#1}}}

\begin{document}

%\date{\today}

\title{Sgluon Pair Production to Next-to-Leading Order}

\author{Dorival Gon\c{c}alves Netto}
\affiliation{Institut f\"ur Theoretische Physik, Universit\"at Heidelberg, Germany}

\author{David L\'opez-Val}
\affiliation{Institut f\"ur Theoretische Physik, Universit\"at Heidelberg, Germany}

\author{Kentarou Mawatari}
\affiliation{Theoretische Natuurkunde and IIHE/ELEM, Vrije Universiteit Brussel, Belgium \\
             and International Solvay Institutes, Brussels, Belgium}

\author{Tilman Plehn}
\affiliation{Institut f\"ur Theoretische Physik, Universit\"at Heidelberg, Germany}

\author{Ioan Wigmore}
\affiliation{SUPA, School of Physics \& Astronomy, The University of Edinburgh, UK}

\begin{abstract}
Scalar color octets are generic signals for new physics at LHC
energies.  We examine their pair production at the LHC to
next-to-leading order QCD. This computation serves as another test of
the fully automized MadGolem framework.  We find large NLO production
rates and sizeable quantum effects which depend on the sgluon mass.
The shift in the sgluon distributions is mild and in good agreement
with a multi-jet merging calculation.
\end{abstract}

\maketitle

%%%%%%%%%%%%%%%%%%%%%%%%%%%%%%%%%%%%%%%%%%%%%%%%%%%%%%%%%%%%%%%%%%%%%%%%
\section{Introduction}
\label{sec:intro}

Sgluons~\cite{sgluons1,sgluons2} are a type of scalar color octet
states which arises in a variety of extensions of the Standard
Model~\cite{review}. They can be fundamental or composite degrees of
freedom.  In extended supersymmetric models like the $R$-symmetric
MSSM~\cite{mrssm,supersoft,sgluons1} or $\mathcal{N}=1/\mathcal{N}=2$
hybrid models~\cite{sgluons2}, sgluons emerge as scalar partners of a
Dirac gluino. More generally, sgluons appear to be ubiquitous in
models of supersymmetry breaking~\cite{susybreaking}. Compositeness
models include fermion fields which transform under a confining gauge
group --- some of them as fundamentals and others as
anti-fundamentals. This naturally leads to scalar states in the $3
\otimes \overline{3} = 1 \otimes 8$ color-adjoint
representations. This mechanism is realized in technicolor and
top-color, chiral-color, and vector-like confinement~\cite{raman}. In
the presence of extra dimensions, color octet scalars emerge as
low-lying Kaluza-Klein modes of the bulk gluon
field~\cite{extradim}.\medskip

At the LHC sgluon pairs will be copiously produced just through their
couplings to gluons. In addition, for large masses, the model
dependent single sgluon production channel might be
competitive~\cite{sgluons1,sgluons2,single_sgluon}.  Available studies
include the single and pairwise production in the context of
supersymmetric scenarios~\cite{sgluons1,sgluons2,phenosusy},
GUTs~\cite{phenogut}, extra dimensions~\cite{extradim,phenoedim}, as
well as more model-independent approaches~\cite{phenomi,manohar_wise}.
Color-octet vector bosons have also been considered ~\cite{vectors_nlo}.
Distinctive decay patterns appear through couplings to pairs of SM
particles or new heavy colored states. The most generic signature is
the decay to two quark or gluon jets, $pp \to \sg\sg^* \to 4$~jets,
possibly including bottom jets~\cite{bottomjets}.  Subjet
techniques~\cite{dijet} or suitable cuts on jet pair invariant
masses~\cite{multijet,exis} have been proposed to handle the
overwhelming QCD background. In supersymmetric models with Dirac
gluinos the constraints on squark mixing are so weak that an
essentially unconstrained squark mass matrix will lead to sgluon
decays to single top (anti-) quarks plus a light jet, $\sg \to
t\bar{q},\bar{t}q$~\cite{sgluons1}. Finally, highly isotropic
multi-jet signatures $pp \to \sg\sg^* \to tt\bar{t}\bar{t} \to 8j +
2\ell + \slashed{E}_T$ are likely for sufficiently heavy
sgluons~\cite{sgluons1,multijet}.  Complementary rare
decays~\cite{raredecays} or long-lived bound
states~\cite{bound_states} are other potential discovery modes for
novel color-adjoint scalars.\medskip

In this paper we present a complete next-to-leading order QCD
calculation of sgluon pair production at the LHC.  We examine the
features and quantitative impact of the QCD quantum effects on the
production rates and sgluon distributions. Our results are implemented
in the {\sc Madgraph} framework through the dedicated {\sc MadGolem}
package for the production of new particles to next-to-leading
order~\cite{madgolemone}.  This tool automatically computes
next-to-leading order QCD corrections for any heavy particle
production process and will be publicly available after the current
testing phase. The NLO sgluon distributions we compare to the
matched~\cite{mlm,lecture} results for the combined process $pp \to
\sg \sg^*$+jets.

Using the renormalizable supersymmetric realization the gluonic QCD
corrections to sgluon pair production are obviously well
defined. Additional supersymmetric QCD corrections are suppressed by
the squark and gluino masses and thus
negligible~\cite{prospino_sqgl,prospino_stop}. Because we are only
interested in sgluon pair production we can decouple all
supersymmetric partners except for the sgluon, retaining all benefits
of a renormalizable theory. This theory can as well be interpreted as
the relevant QCD part of an effective strongly interacting theory.

%%%%%%%%%%%%%%%%%%%%%%%%%%%%%%%%%%%%%%%%%%%%%%%%%%%%%%%%%%%%
\section{Sgluon pair production to NLO}

\begin{figure}[t]
\includegraphics[width=0.15\textwidth]{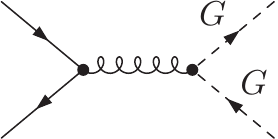} 
\hspace*{0.1\textwidth}
\includegraphics[width=0.6\textwidth]{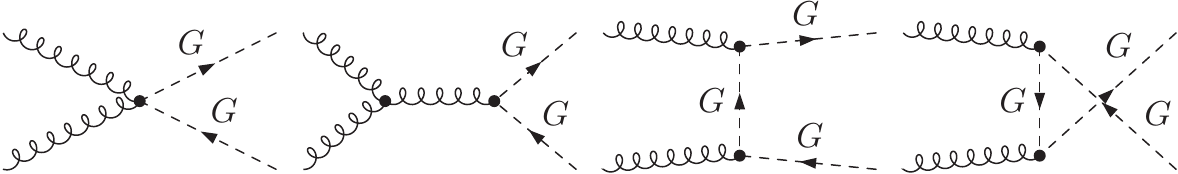}
\caption{Leading order Feynman diagrams for sgluon pair production via
  quark-antiquark annihilation and gluon fusion.}
\label{fig:born}
\end{figure}

To compute the complete NLO corrections for sgluon pair production at
the LHC we minimally extend the Standard Model by one additional color
octet, weak singlet, electrically neutral, and complex scalar
field $\sg$.  The sgluon couples to the Standard Model through the
covariant derivative, $D_\mu\, G^A \equiv \partial_\mu\, G^A +
g_s\,f^{ABC}\,G^B\, A_\mu^C$, where $A_\mu^C$ denotes the gluon field,
$g_s$ the strong coupling constant, and $f^{ABC}$ the adjoint $SU(3)$
generators.  The Feynman diagrams for the two partonic LHC production
mechanisms
\begin{equation}
q\bar{q} \to \sg \sg^* 
\qquad \text{and} \qquad  
gg \to \sg \sg^*
\end{equation}
are shown in Figure~\ref{fig:born}.  The sgluon coupling to gluons
reads
\begin{alignat}{5}
\lag 
\supset&  \; D_\mu \sg^* \; D^\mu \sg -m_\sg^2 \sg\sg^* 
\notag \\
\supset& \;
-g_s\, f^{ABC}\,  \left[\sg^{A*}(\partial^\mu\,\sg^B) - (\partial^\mu\,\sg^{A*})\sg^B\, 
                \right] \,A_\mu^C
+ g_s^2\, \left[ f^{ACE}\,f^{BDE} + f^{ADE}\,f^{BCE} \right]
\; \sg^{C*}\,\sg^D A^A_\mu\,A^{B \mu} \; .
\label{eq:lagone}
\end{alignat}
Incidentally, we notice the absence of direct sgluon couplings to
matter. In supersymmetry they only arise as effective dimension-5
operators induced by the one-loop squark and gluino loops with a
non-trivial scaling for individual heavy masses~\cite{sgluons1}.  As
long as these couplings are small --- which is true if they are
loop-induced --- the sgluon mass range is not constrained by stringent
bounds from dijet resonance searches. As a consequence, sgluons can be
relatively light.  Conversely, for $\mathcal{O}(1)$ sgluon-quark-quark
couplings sgluon masses below $m_\sg = \mathcal{O}(2~\tev)$ are ruled
out by the LHC experiments.\medskip

%------------------------------------------------
%------------------------------------------------

We implement the couplings shown in Eq.\eqref{eq:lagone} into the {\sc
  Madgraph} framework~\cite{madgraph}. {\sc MadGolem} then generates
all tree-level diagrams and the corresponding helicity amplitudes,
making use of {\sc Madgraph} and {\sc Helas}~\cite{helas}.  All
one-loop amplitudes and the corresponding helicity amplitudes we
generate with a modified version of {\sc Qgraf}~\cite{qgraf} and {\sc
  Golem}~\cite{golem,golem_lib}. The model specific ultraviolet
counter terms are part of the model implementation.  The subtraction
of infrared and (if applicable) on-shell divergences is
automized~\cite{prospino_sqgl,prospino_onshell}.

Throughout our analysis we use the CTEQ6L1 and CTEQ6M parton
densities~\cite{cteq} with consistent values of $\alpha_s$.  For the
central renormalization and factorization scales we choose the average
final state mass $\mu^0 \equiv \mu_{R,F} = m_G$, which has been shown to lead to stable perturbative
results~\cite{prospino_sqgl,prospino_stop}. The LHC center of mass
energy is $\sqrt{S} = 8$~TeV. Unless stated otherwise, we set the
sgluon mass to $m_\sg = 500$~GeV.\medskip

Technically, there is a {\sc MadGraph4}~\cite{madgraph} issue with the
color structure of the quartic gluon-gluon-sgluon-sgluon coupling
shown in Eq.\eqref{eq:lagone}. We therefore generate the required
structure $f^{ACE}\,f^{BDE} + f^{ADE}\,f^{BCE}$ through an auxiliary
massive, color-adjoint vector boson $V_\mu$ with an appropriate
coupling to a gluon and a sgluon, namely
\begin{equation}
\sg^{A*}A^B_\mu\,V^{C \mu}: \;  g_s\,m_V\,f^{ABC}
\qqquad
\sg^A \,A^B_\mu\,V^{C \mu}: \; -g_s\,m_V\,f^{ABC} \; .
\label{eq:xfield}
\end{equation}
The quartic gluon-sgluon interaction is then simply given by the
decoupling limit $m^2_V \gg s$. For {\sc Madgraph5} this technical
complication is not necessary any longer.

%%%%%%%%%%%%%%%%%%%%%%%%%%%%%%%%%%%%%%%%%%%%%%%%%%%%%%%%%%%%%%%%%%%%%%%%
\subsection*{Production rates to Next-to-Leading order}

As a first step we present the results for the total NLO cross section
for sgluon pair production. Later in this section we focus on more
specific aspects of the real and virtual corrections.  Unless stated
otherwise, we assume $m_\sg = 500$~GeV and $\sqrt{S} = 8$~TeV.

The size of the QCD quantum effects we describe in terms of the
consistent factor $K \equiv \sigma^\text{NLO}/\sigma^\text{LO}$. From
the production of supersymmetric particles it is well known that for
LHC energies of 8~TeV and particle masses in the 500~GeV to 1~TeV mass
range this correction factor can become unexpectedly large. This is
not a sign of poor perturbative behavior but an artifact of the LO
CTEQ parton densities~\cite{prospino_private,cteq}. Correspondingly,
Table~\ref{tab:xsec} typically shows $K \gtrsim 1.5$ for this collider
energy while the 14~TeV scenario has smaller, yet sizeable,
corrections.\medskip

%------------------------------------------------
\begin{table}[t]
\begin{tabular}{c||l|l|c||l|l|c} \hline
 & \multicolumn{3}{c||}{$\sqrt{S} = 8\,\tev$}  & \multicolumn{3}{c}{$\sqrt{S} = 14\,\tev$} \\ \hline
 $m_\sg$ [GeV]  & $\sigma^\text{LO} [\text{pb}]$ & $\sigma^\text{NLO} [\text{pb}]$ & $K$ &  $\sigma^\text{LO} [\text{pb}]$ & $\sigma^\text{NLO} [\text{pb}]$ & $K$ \\ \hline\hline
 200 & $2.12 \times 10^2$ & $3.36\times 10^2$ & 1.58 & $9.77\times 10^2$ & $1.48\times 10^{3}$ & 1.52   \\
 350 & $8.16\times 10^{0}$ & $1.36\times 10^{1}$ & 1.66 & $5.44\times 10^1$ & $8.46\times10^{1}$ & 1.56 \\
 500 & $7.64\times10^{-1}$ & $1.34\times10^{0}$ & 1.75 & $7.14\times10^{0}$ & $1.14\times10^1$ & 1.60 \\
 750 & $3.40\times10^{-2}$ & $6.54\times10^{-2}$ & 1.93 & $5.56\times10^{-1}$ & $9.29\times10^{-1}$ & 1.67\\
 1000 & $2.47\times 10^{-3}$ & $5.29\times 10^{-3}$ & 2.15 & $7.31\times 10^{-2}$ & $1.28\times 10^{-1}$& 1.75 \\ \hline
\end{tabular}
\caption{Total $pp \to \sg \sg^*$ cross sections and corresponding
  $K$-factors for different sgluon masses and LHC energies.}
\label{tab:xsec}
\end{table}
%------------------------------------------------

%------------------------------------------------
\begin{figure}[b]
\includegraphics[width=0.3\textwidth, angle=-90]{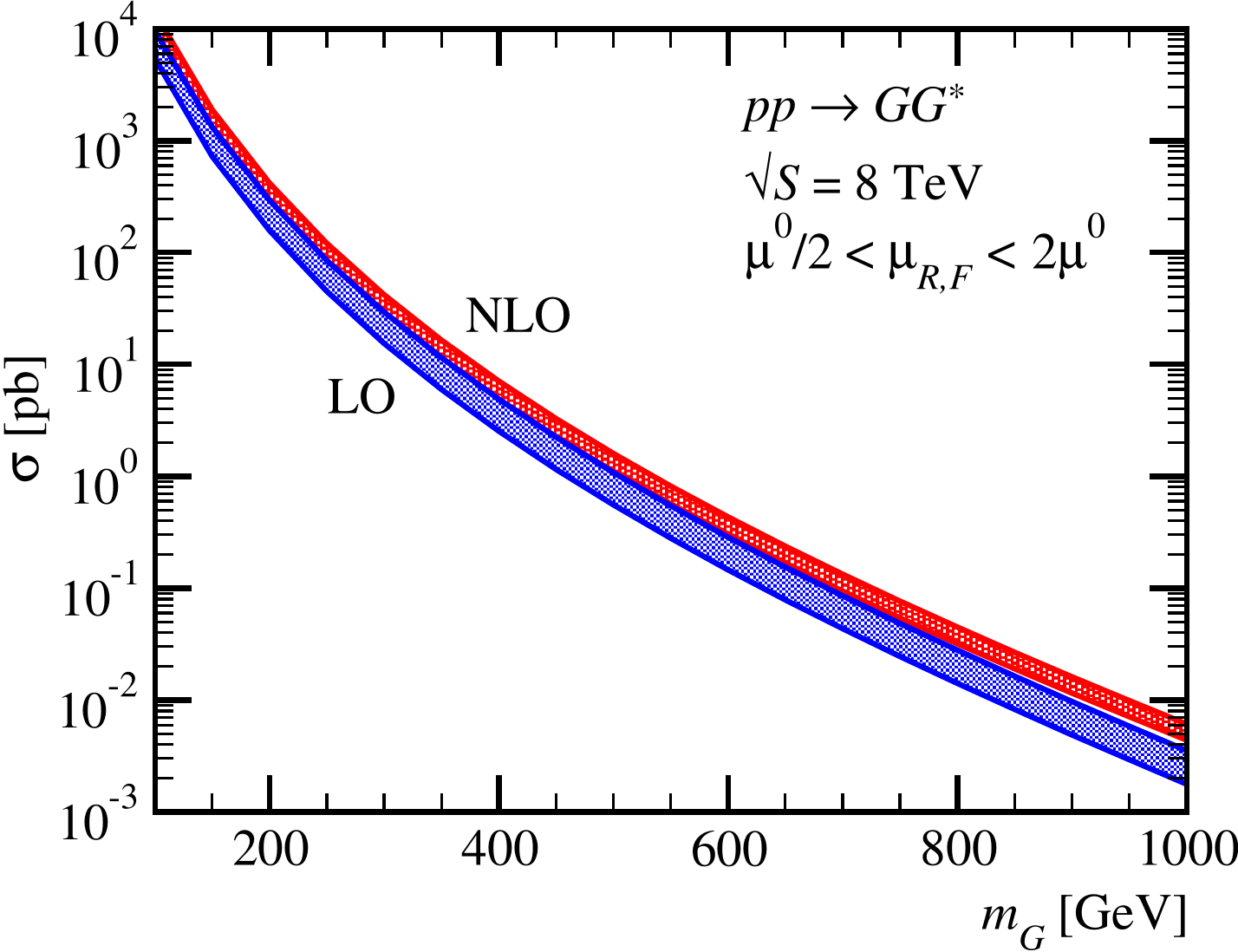}
\hspace*{0.1\textwidth}
\includegraphics[width=0.3\textwidth, angle=-90]{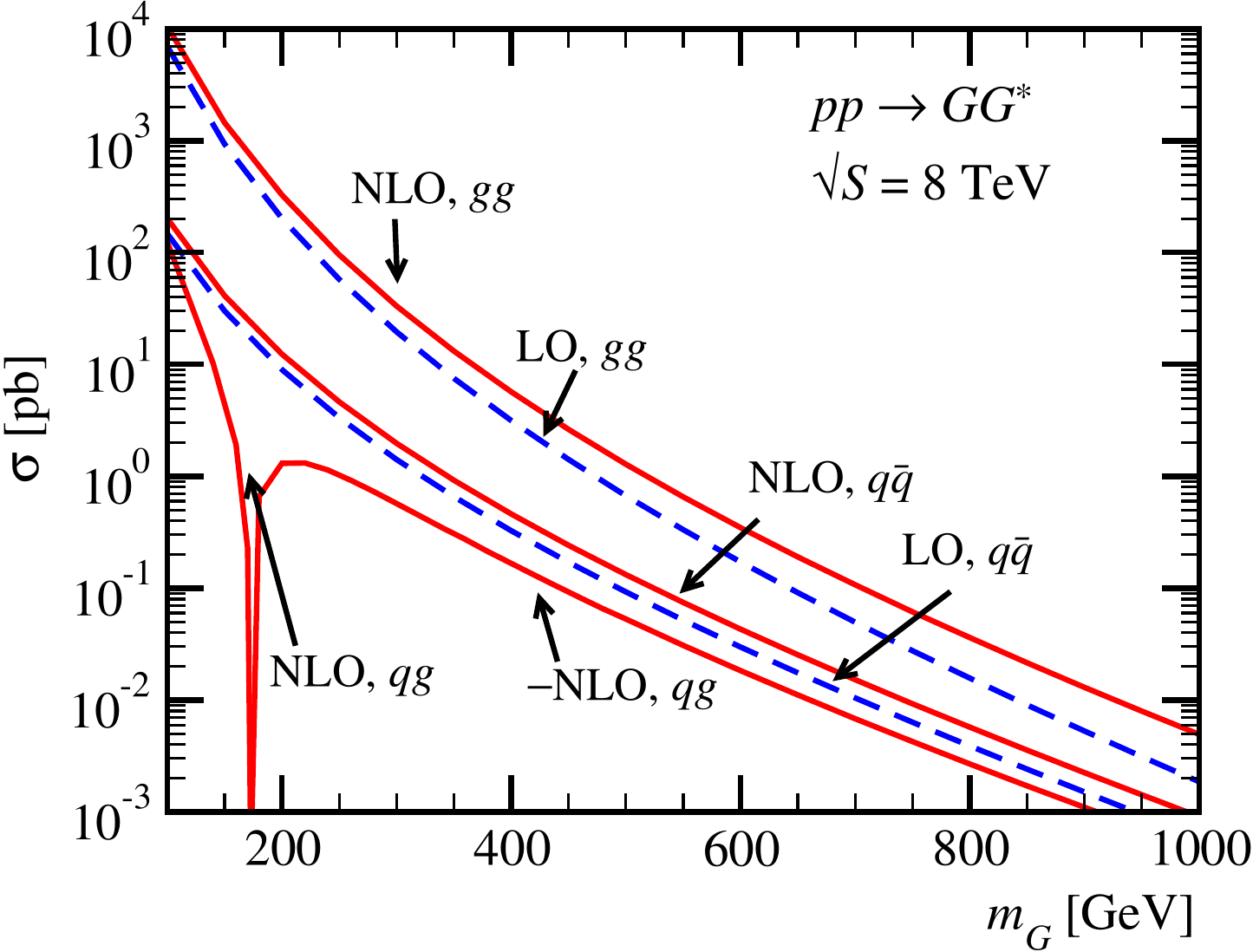}
\caption{LO and NLO cross sections $\sigma(pp \to \sg\sg^*)$ as a
  function of the sgluon mass. The band corresponds to a scale variation
  $\mu^0/2 < \mu_{R,F} < 2\mu^0$. In the right panels we explicitly
  separate the contributions from the different partonic sub-channels,
  $q\bar{q}$, $gg$ and also $gq$.}
\label{fig:overmass}
\end{figure}
%------------------------------------------------

\begin{figure}[t]
\includegraphics[width=0.8\textwidth]{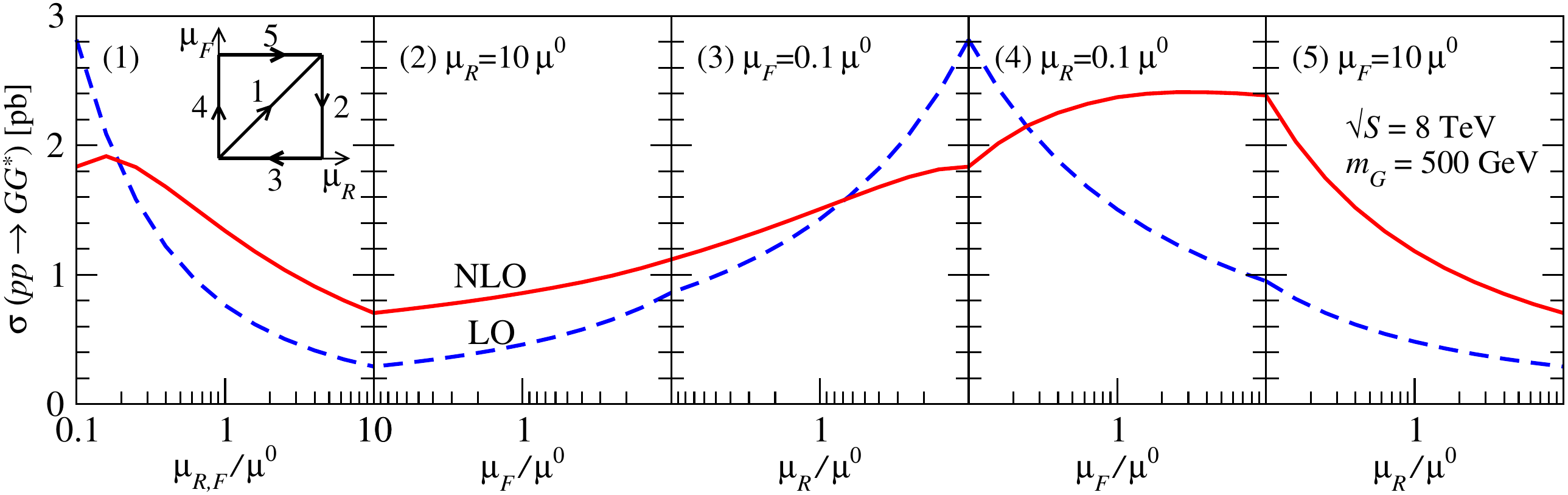} 
\caption{Renormalization and factorization scale dependence. The plot
  traces the scale dependence following a contour in the
  $\mu_F$-$\mu_R$ plane in the range $\mu = (0.1 - 10) \times \mu^{0}$
  with $\mu^{0} = m_\sg$.  The sgluon mass we fix to $m_\sg = 500$ GeV.}
\label{fig:scale}
\end{figure}

We then provide a comprehensive analysis of the LO and NLO cross
sections $\sigma(pp \to \sg\sg^*)$ as a function of the sgluon mass in
Fig.~\ref{fig:overmass}. In the left panel we show the LO and NLO
cross sections with the envelope of the NLO factorization and
renormalization scale variation in the range $\mu^0/2 < \mu_{R,F} < 2\mu^0$. 
The effects of an independent as well
as diagonal variation of the factorization and renormalization scales
we show in Fig.~\ref{fig:scale} and discuss below. As alluded to, the
LO parton densities drive the LO cross sections to unexpectedly small
values which makes the NLO corrections appear larger than $\sim 100
\%$ for sgluon masses in the TeV range.

In the right panel of Fig.~\ref{fig:overmass} we separate the
contributions stemming from the different partonic sub-channels:
$q\bar{q}$, $gg$ and the crossed purely NLO $gq/g\bar{q}$ initial
state.  The NLO corrections steadily increase for increasing sgluon
masses.  In part, we can trace back this behavior to threshold effects
which we will further discuss below. The right panel of
Fig.~\ref{fig:overmass} also quantifies the dominance of the
gluon-fusion mechanism $gg \to \sg\sg^*$.  The reason is twofold:
first, the color charges in the four-octet interaction are larger than
the triplet-octet combination that drives the $q\bar{q}$ channel.
Second, the $gg$ channel benefits from particular kinematic features
of the parton level process.  While $q\bar{q} \to \sg\sg^*$ to LO
proceeds merely through the (derivative) $g\sg\sg$ coupling, the $gg$
mechanism also receives a contribution from the (contact) quartic
interaction.  The first case corresponds to a $p$-wave and implies
that the total (partonic) cross section scales as $\sigma_{q\bar{q}}
\sim \beta^3$ at threshold, where $\beta = \sqrt{1-4m_\sg^2/s}$
denotes the sgluon velocity in the center-of-mass frame. In the gluon
fusion case the $s$-wave component from the quartic interaction
translates into a linear dependence $\sigma(gg) \sim \beta$. The
latter dominates in the vicinity of the threshold which, moreover,
corresponds to the low-$x$ region where the gluon parton densities
becomes large.\medskip

%------------------------------------------------
%------------------------------------------------

Following these arguments we can compare sgluon pair production to
stop pair production (or squark pair production with decoupled
gluinos)~\cite{prospino_stop}. The differences at leading order can
be traced to the relative strength of the color interactions arising
from the fundamental vs adjoint final-state scalars. The ratios of the
expected sgluon versus stop pair production rates can be roughly
inferred from their parton level cross sections. We can
compute these ratios directly from the corresponding analytical
expressions~\cite{sgluons1,sgluons2},
\begin{equation}
\frac{\sigma(q\bar{q} \to \tilde{t}\tilde{t}^*)}
     {\sigma(q\bar{q} \to \sg\sg^*)} = 1/6 
\qquad \qquad 
\frac{\sigma(gg \to \tilde{t}\tilde{t}^*)}
     {\sigma(gg \to \sg\sg^*)} \simeq 1/20 \;  .
\label{eq:sqvssg}
\end{equation}
These estimates nicely agree with the NLO results for stop pair
production available from {\sc Prospino}~\cite{prospino_stop}, which
give $\sigma(pp \to \tilde{t}\tilde{t}^*) \sim 3 $~pb for stop masses
of $m_{\tilde{t}} \sim 350\,\gev$, \ie a factor of $\mathcal{O}(20)$
below the sgluon results in Table~\ref{tab:xsec} and
Fig.~\ref{fig:overmass}. The NLO effects to squark pair production are
comparatively mild. In contrast, gluino pair production as an example
of a color-octet interaction also shows large $K$ factors and a very
pronounced dependence on the mass of the produced heavy particles.

%%%%%%%%%%%%%%%%%%%%%%%%%%%%%%%%%%%%%%%%%%%%%%%%%%%%%%%%%%%%%%%%%%%%%%%%
\subsection*{Scale dependence}

Aside from the often positive corrections to the production rate the
main motivation for the computation of higher order corrections is the
reduced theoretical uncertainty. While we cannot derive the
uncertainties arising from unknown higher orders in QCD from first
principles, we can attempt to track them in the dependence of
unphysical parameters introduced by the perturbative approach. An
example for such a parameter are the factorization and renormalization
scales which we introduce when we remove collinear and ultraviolet
divergences order by order in perturbation theory. In the limit of
infinitely many terms in the power series in $\alpha_s$ these scale
dependences have to vanish, so unless there is a systematic shift from
one perturbative order to the next the scale dependence should cover
the asymptotic cross section values.  While for Drell-Yan-type
processes we know that this argument fails, purely color mediated
processes with $\sigma^\text{LO} \propto \alpha_s^2$ tend to give a
reasonable error estimate this way~\cite{prospino_sqgl,prospino_stop}.
Vice versa, we can at least firmly state that the scale variation
gives a minimum uncertainty simply because we have the freedom to
choose the two scales within a reasonable energy range.\medskip

In Fig.~\ref{fig:scale} we show the scale dependence for the $pp \to
\sg\sg^*$ production rate, independently changing the renormalization
($\mu_R$) and the factorization ($\mu_F$) scales. We illustrate the
path in two dimensions in the little square in the first panel.  The
individual scale variation is chosen as $\mu^{(0)}/10 < \mu <
10\mu^{(0)}$, where $\mu^{(0)}$ stands for our central value choice
$\mu_F = \mu_R = \mu^{(0)} = m_\sg = 500$~GeV.  The stabilization of
the mentioned scale dependence manifests itself as a smoother
$\sigma^\text{NLO}$ slope, with varies typically around $\Delta
\sigma^\text{NLO}/\sigma^\text{NLO} \sim \mathcal{O}(30\%)$, while for
leading order this variation can be as large as $\mathcal{O}(80\%)$.

We also see that the maximum rate at small scale values which is often
interpreted as a sign of scale stabilization is an artifact of
identifying the two scales.  An independent variation gives the
largest rates at small values of the renormalization scale combined
with larger values of the factorization scale --- even though from a
resummation point of view it is not clear how such a scale choice
would be interpreted~\cite{lecture}.

%%%%%%%%%%%%%%%%%%%%%%%%%%%%%%%%%%%%%%%%%%%%%%%%%%%%%%%%%%%%%%%%%%%%%%%%
\subsection*{Real emission}

\begin{figure}[t]
\begin{center}
 \includegraphics[width=0.6\textwidth]{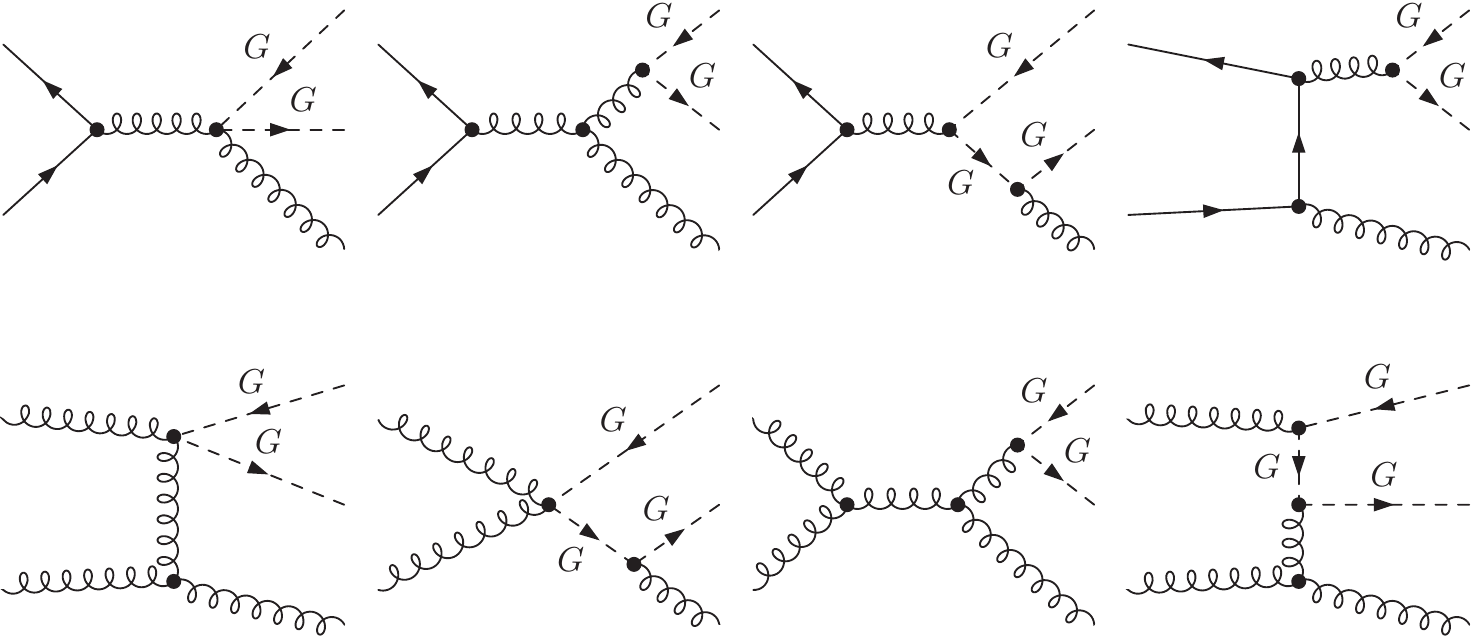} 
\end{center}
\caption{Sample Feynman diagrams for real emission corrections to
  sgluon pair production via quark-antiquark annihilation (upper) and
  gluon fusion (lower).}
\label{fig:real}
\end{figure}

Real emission corrections to sgluon pair production arise at order
$\alpha_s^3$ and originate from the three-particle final state
contributions, wherein one extra gluon accompanies the produced sgluon
pair. We show sample Feynman diagrams in Fig.~\ref{fig:real}.
Following the standard procedure we subtract infrared divergences from
the emitted gluon in the soft and/or collinear regimes using the
massive Catani-Seymour
dipoles~\cite{catani_seymour_massless,catani_seymour_massive}. In
addition to the SM dipoles available in the {\sc
  MadDipole}~\cite{maddipole} package, {\sc MadGolem} includes for
example the sgluon dipoles to cope with our novel infrared divergent
structure. Such divergences appear when the external sgluons radiate
soft gluons and require new final-final and final-initial dipoles.
The sgluon can also be a heavy spectator parton, but for this case we
can simply use the SM initial-final dipoles. This is because for the
dipole function the spectator carries information about the mass of
the colored particle, but not about its spin.

%------------------------------------------------
%------------------------------------------------

In Appendix~\ref{sec:dipoles} we give the new sgluon dipoles including
the FKS-style phase space cutoff $0<\alpha \leq 1$~\cite{alpha}. The
numerical implementation is publicly available upon request. Among
several numerical improvements, the parameter $\alpha$ gives us an
easy handle to check our implementation. For a wide range $\alpha =
10^0 - 10^{-8}$ we find stable cross section for the combination of
the real emission diagrams with the integrated dipoles. As a default
value in {\sc MadGolem} we use $\alpha = 10^{-3}$.

%%%%%%%%%%%%%%%%%%%%%%%%%%%%%%%%%%%%%%%%%%%%%%%%%%%%%%%%%%%%%%%%%%%%%%%%
\subsection*{Virtual corrections}

%------------------------------------------------
\begin{figure}[b]
\begin{center}
 \includegraphics[width=0.6\textwidth]{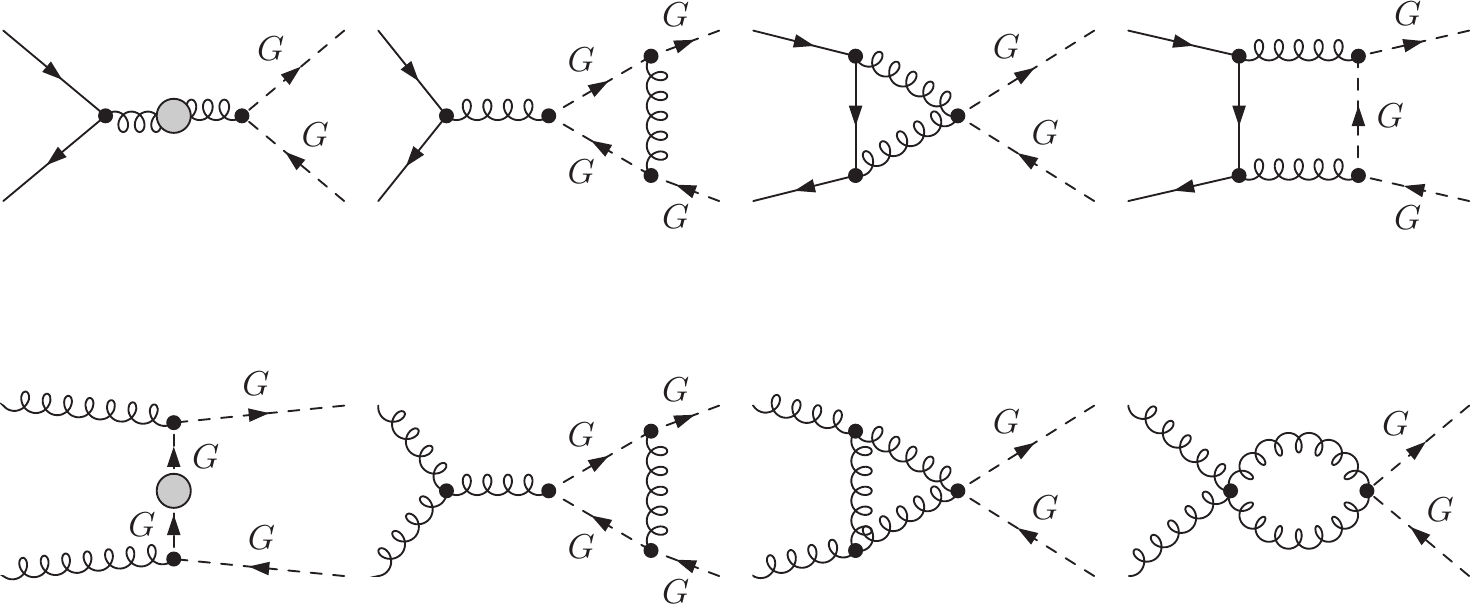} 
\end{center}
\caption{Sample Feynman diagrams for virtual corrections to sgluon
  pair production via quark-antiquark annihilation (upper) and gluon
  fusion (lower).}
\label{fig:virtual}
\end{figure}
%------------------------------------------------

%------------------------------------------------
\begin{figure}[t]
\begin{center}
\includegraphics[width=0.3\textwidth, angle=-90]{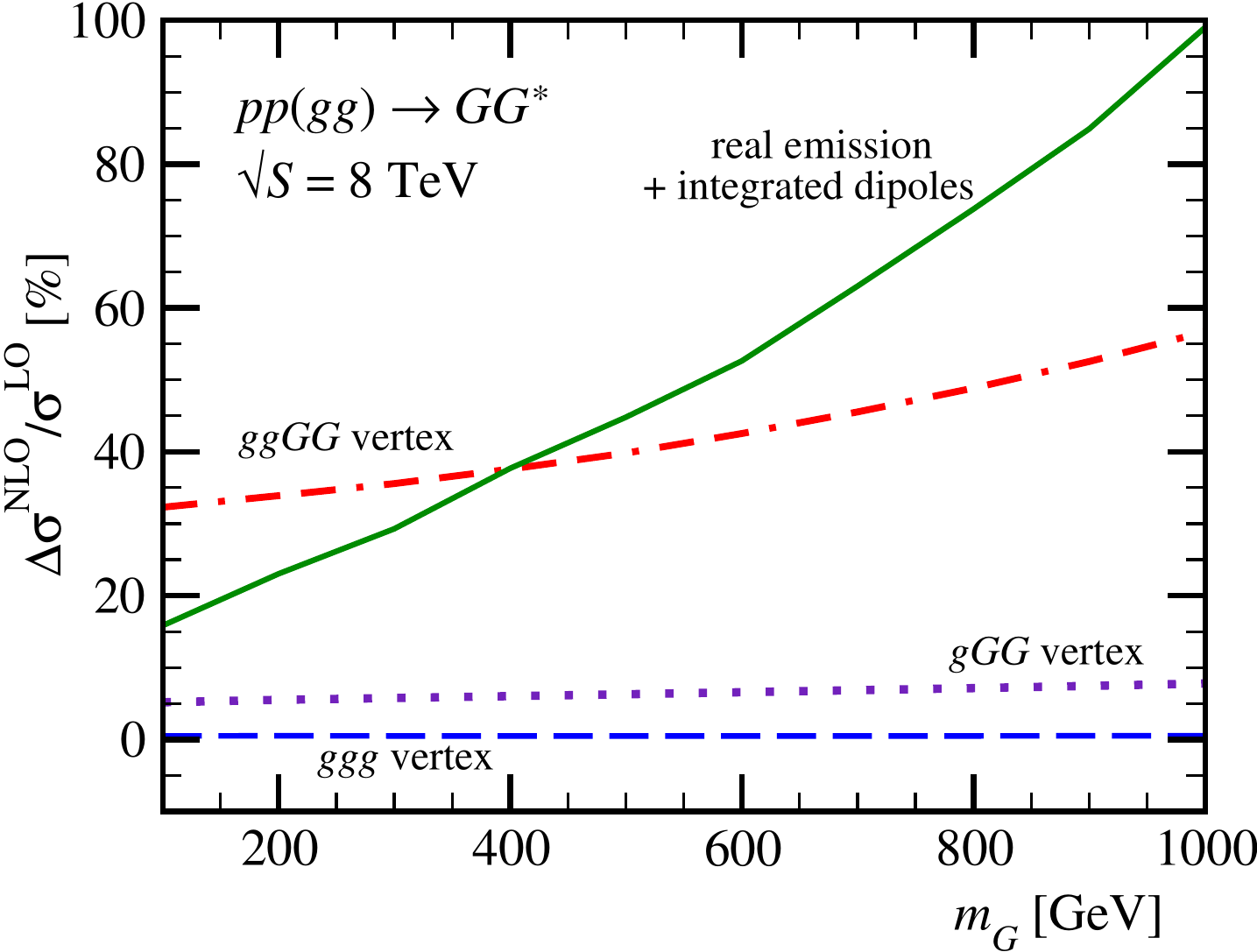}
\hspace*{0.1\textwidth}
\includegraphics[width=0.3\textwidth, angle=-90]{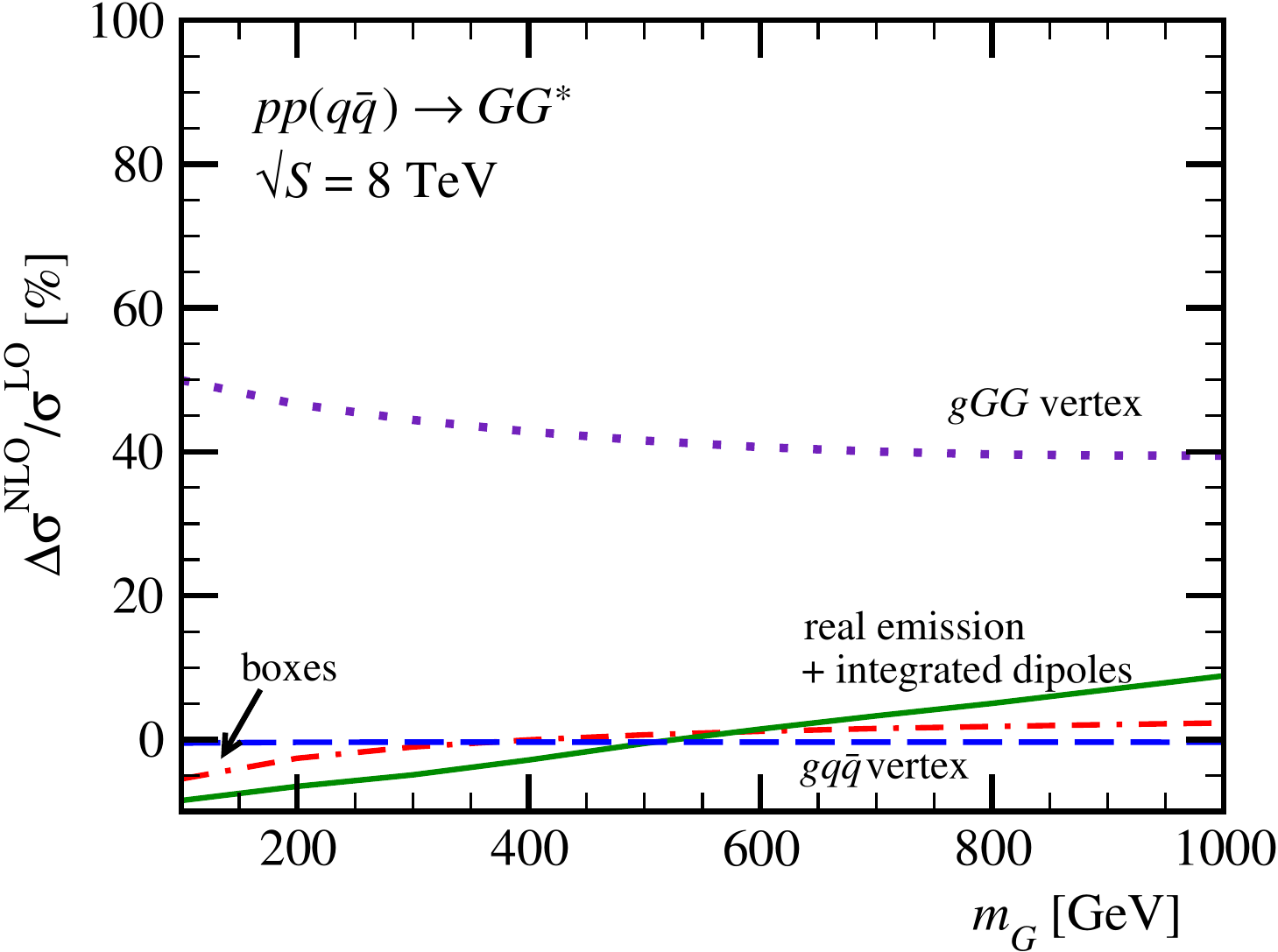}
\end{center}
\caption{Relative size $\Delta\sigma^\text{NLO}/\sigma^\text{LO}
  \equiv (\sigma^\text{NLO} - \sigma^\text{LO})/\sigma^\text{LO}$ of
  the real emission and virtual corrections to $\sigma(pp \to
  \sg\sg^*)$ as a function of the sgluon mass $m_\sg$. We separate the
  partonic $gg$ (left) and $q\bar{q}$ (right) initial states. The contribution
from the self-energies is negligible and not explicitly shown.}
\label{fig:virtuals}
\end{figure}
%------------------------------------------------ 

Virtual corrections to sgluon pair production appear as order
$\alpha_s^3$ contributions from virtual gluons coupling to quarks and
sgluons. All divergences we regularize in $n = 4-2\epsilon$
dimensions. The infrared poles are cancelled after we include the
integrated Catani-Seymour
dipoles~\cite{catani_seymour_massless,catani_seymour_massive} and take
into account the collinear higher order correction consistently
included in the definition of the parton densities.  The ultraviolet
divergences are absorbed in the physical renormalization of the strong
coupling constant and the sgluon mass. As described in
Appendix~\ref{sec:UV-renorm} we use the $\msbar$ scheme with decoupled
heavy colored states~\cite{decoupling} for the strong coupling and the
on-shell scheme for the mass. For an internal check we use an
independent implementation of our sgluon model in {\sc FeynArts}; all
{\sc MadGolem} results can then be numerically compared to the output
from {\sc FeynArts, FormCalc} and {\sc
  LoopTools}~\cite{feynarts}.\medskip

Starting with the dominant gluon fusion channel, in the left panel of
Fig.~\ref{fig:virtuals} we examine different contributions to the real
and virtual NLO corrections to the hadronic process $pp \to \sg\sg^*$
as a function of the sgluon mass. Leaving aside gauge invariance
issues while applying a numerical test we separately show different
one-loop pieces normalized to the LO rate,
$\Delta\sigma^\text{NLO}/\sigma^\text{LO}$. In addition, we
distinguish the partonic subprocesses $q\bar{q}$ and $gg$. The crossed
channel $qg$ does not develop any virtual corrections but is required
for the complete cancellation of the collinear divergence.

The real emission together with the virtual box diagrams contributes
the bulk of the NLO quantum effects. Both feature a characteristic
growing trend with increasing sgluon mass. For intermediate sgluon
masses, real emission gives rise to corrections in the ball-park of
$20-60\%$, but it may eventually reach up to $100\%$ for TeV-scale
sgluons. Note that this behavior cannot be interpreted as a break-down
of perturbation theory because QCD corrections to our supersymmetric
setup are fundamentally well defined.  Box-like topologies, \ie
one-loop corrections to the $ggGG$ vertex including diagrams shown as
the 3th and 4th diagrams in the lower row of Fig.~\ref{fig:virtual},
amount to roughly $40\%$ and exhibit a slightly milder dependence on
$m_\sg$. Both, the size of these contributions and their increase with
$m_\sg$ we can attribute to the peculiar threshold behavior of the NLO
corrections.  In particular, long-range gluon exchange between
slowly moving heavy final-state particles $\beta \to 0$ develops a
Coulomb singularity $\sigma^\text{NLO} \sim \pi\alpha_s/\beta$ which
cancels the linear dependence from the tree-level contribution
$\sigma^\text{LO} \sim \beta$ and leads to a finite NLO rate but a
divergent $K$ factor~\cite{prospino_stop}.  Gluon radiation off the
initial state carries positive, and potentially large, logarithmic
pieces which supply an additional source of enhancement -- and that
can eventually be resummed~\cite{single_sgluon}.\medskip

For the subleading $q\bar{q}$ initial state we find a sizable
contribution from the gluon-sgluon-sgluon $g\sg\sg$ vertex
corrections. They are fairly independent of variations of the sgluon
mass. The remaining one-loop topologies only contribute at the percent
level of less.

%%%%%%%%%%%%%%%%%%%%%%%%%%%%%%%%%%%%%%%%%%%%%%%%%%%%%%%%%%%%%%%%%%%%%%%%
\section{Distributions: NLO versus multi-jet merging}
\label{sec:merging}

Predictions based on (fixed-order) NLO cross sections entail
significant improvements of the central values and the theory
uncertainties, as we have just shown for the specific case $pp \to
\sg\sg^*$. Before these results can be integrated in experimental
analyses we need to confirm that this quantitative picture also holds
for the main distributions. Previous work in the literature shows that
the transverse momentum and rapidity distributions of pair produced
heavy particles are relatively stable with respect to higher-order
corrections~\cite{prospino_sqgl,madgolemone}.  For the production of
such heavy particles a parton shower should deliver a good description
of jet radiation patterns, because its underlying collinear
approximation applies for a wide range of transverse momenta relative
to the sgluon
masses~\cite{skands,sgluons1,madgraph_merging}. Normalizing the event
numbers generated by standard Monte Carlo tools to the NLO cross
section should therefore be an appropriate strategy.\medskip

\begin{figure}[t]
\begin{center}
\includegraphics[width=0.35\textwidth, angle=-90]{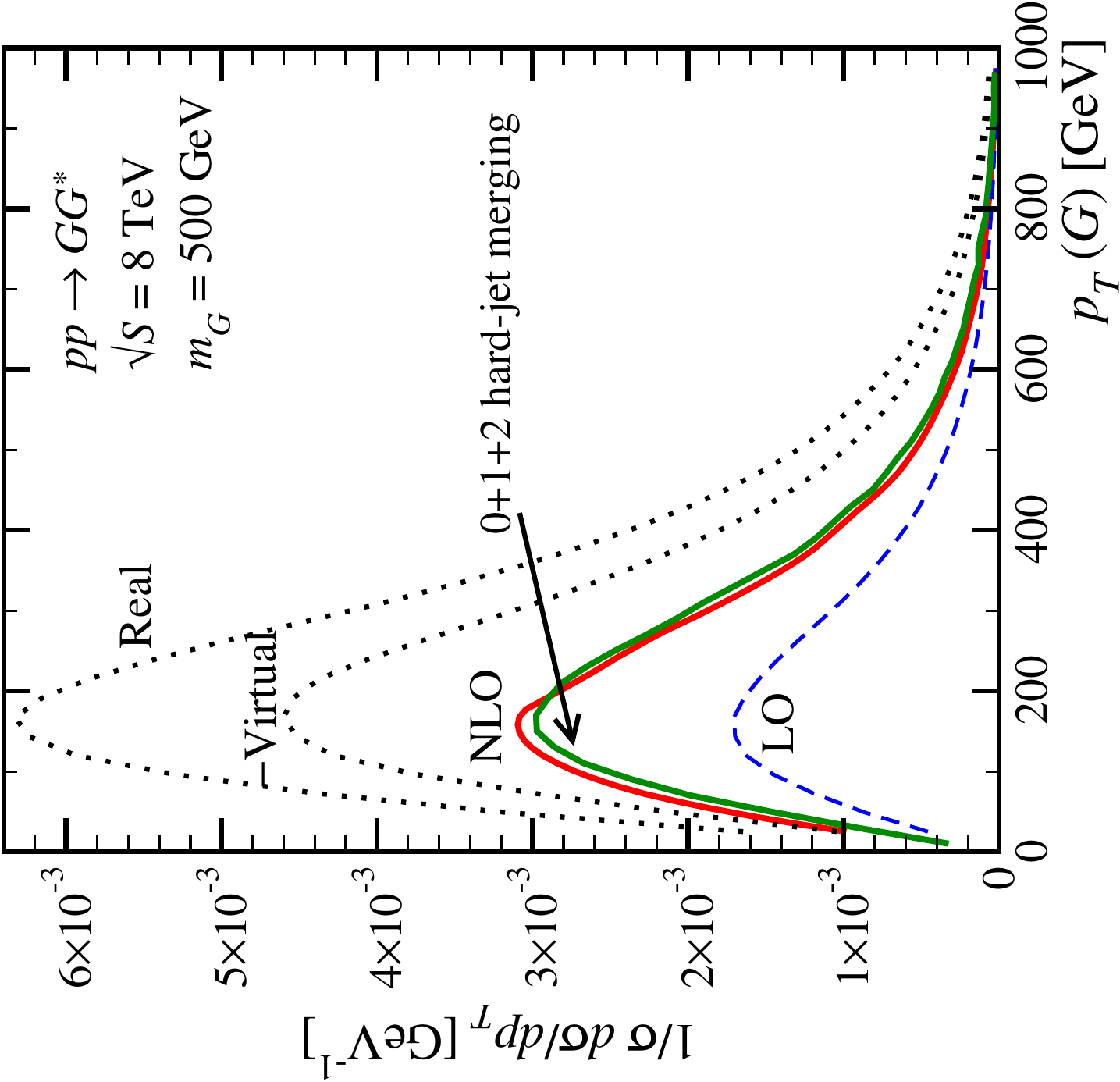}
\hspace*{0.1\textwidth}
\includegraphics[width=0.35\textwidth, angle=-90]{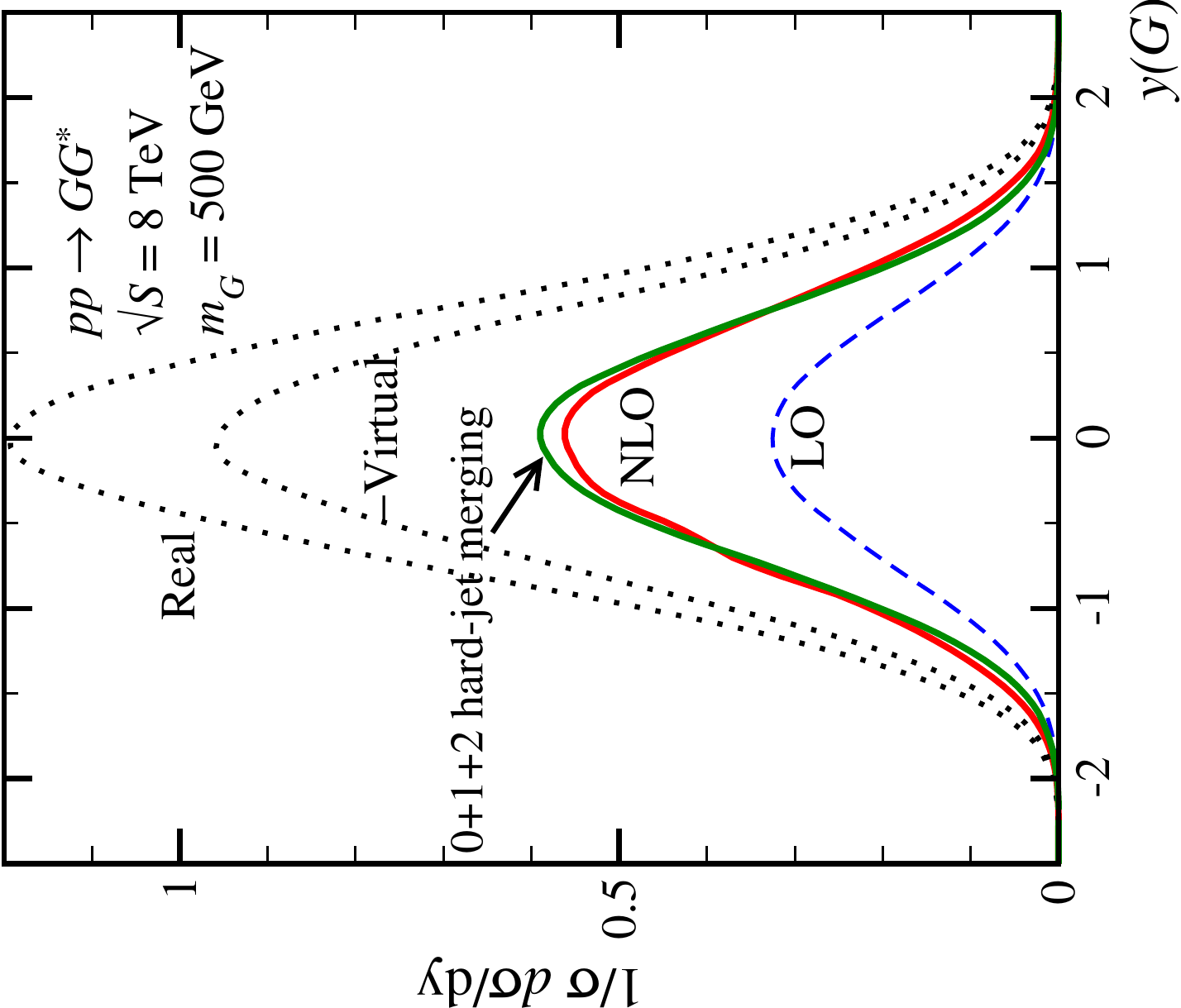}
\end{center}
\caption{Sgluon transverse momentum and rapidity distributions at
  parton level. We assume $m_\sg = 500$~GeV and $\sqrt{S} = 8$~TeV.
  For the NLO curves we separately display the LO, virtual, and real
  contributions ($\alpha = 10^{-3}$).  In addition, we show the
  corresponding distributions based on multi-jet merging in the MLM
  scheme~\cite{mlm} with up to two hard radiation jets. The NLO and
  merged results are normalized to unity while the different
  contributions to the NLO rates are shown to scale.}
\label{fig:distrib}
\end{figure}

To quantitatively assess such statements we compare the fixed order
NLO parton-level distributions for the production process $pp \to
\sg\sg^*$, as obtained from {\sc MadGolem}, with a multi-jet merging
calculation. For the latter we employ the {\sc Mlm}~\cite{mlm} scheme
and generate events using {\sc MadGraph 4.5}~\cite{madgraph_merging}
interfaced with {\sc Pythia}~\cite{pythia}. The entire description of
our sgluon is supplied by the {\sc MadGolem} model file. In
Fig.~\ref{fig:distrib} we display the resulting transverse momentum
and rapidity distributions for one outgoing sgluon for the NLO
calculations as well as for jet merging including up to two hard
jets. Only including one hard jet in the merging prescription would
not change the results within their numerical precision.  Those two
distributions are normalized to unity.  For the NLO results we
separately show the LO, real emission, and virtual gluon
contributions, defined in terms of Catani-Seymour dipoles with $\alpha
= 10^{-3}$.\medskip

%------------------------------------------------
%------------------------------------------------

First, we see that all different ingredients of the NLO distributions
have essentially the same shapes. Large effects on the total rate from
collinear radiation or Coulomb singularities only have a negligible
effect on the distributions of the heavy states.  In addition, the
normalized fixed-order and merged distributions agree very well. Small
differences like the slightly harder $p_T$ profile of the merged
prediction are accounted for by the extra recoil jets from the parton
shower. Similarly, such a second jet from initial state radiation can
balance the first emission and lead to more central sgluons in the
detector.

%%%%%%%%%%%%%%%%%%%%%%%%%%%%%%%%%%%%%%%%%%%%%%%%%%%%%%%%%%%%%%%%%%%%%%%%
\section{Summary}
\label{sec:sum}

We report on the first complete calculation of sgluon pair production
to next-to-leading order. The pairwise production of scalar
color-adjoints we define in terms of an extended supersymmetric model;
however, after decoupling squarks and gluinos our results can be
considered reasonably model-independent.  The sgluons have tree level
QCD couplings to gluons and do not couple to matter. We find
\begin{enumerate}
\item potentially large production rates, driven by the gluon fusion
  subprocess. Typical numbers range around $\sigma \sim
  \mathcal{O}(1~\pb)$ for sgluon masses of $m_\sg \simeq 500$~GeV and
  $\sqrt{S} = 8\,\tev$. Such small masses are not ruled out by current
  experimental constraints.
\item sizable NLO quantum effects, traced back to real gluon radiation
  and a certain subsets of vertex and box virtual corrections.  Their
  relative size increases with the sgluon mass, mainly due to
  threshold effects.
\item substantially reduced theoretical uncertainties.  The scale
  dependence which is dominated by the renormalization scale choice
  and which may reach $\mathcal{O}(80\%)$ at leading order, is reduced
  by a factor $1/2-1/4$.
\item NLO sgluon distributions which agree very well with
  complementary results from multi-jet merging.  Applying the NLO rate
  normalization to sgluons+jets event generation should give very
  consistent predictions for the LHC.
\end{enumerate}

Besides its phenomenological impact our study illustrates the
performance of the (soon-to-be-public) {\sc MadGolem} package. The
genuine dipole and counter term structures which cope with
infrared and ultraviolet divergences in the presence of the sgluon
field have been implemented and can be accessed automatically.  In
this sense the present study qualifies as a non-trivial example of a
fully automized NLO calculation for the production of heavy particles
beyond the Standard Model.

\acknowledgments

KM and IW would like to thank the Institute for Theoretical Physics
and Heidelberg University for their support and hospitality during
many visits. DG acknowledges support by the International Max Planck
Research School for Precision Tests of Fundamental Symmetries. This
work is in part supported by the Concerted Research action
``Supersymmetric Models and their Signatures at the Large Hadron
Collider'' of the Vrije Universiteit Brussel and by the Belgian
Federal Science Policy Office through the Interuniversity Attraction
Pole IAP VI/11.

\appendix

%%%%%%%%%%%%%%%%%%%%%%%%%%%%%%%%%%%%%%%%%%%%%%%%%%%%%%%%%%%%%%%%%%%%%%%%
\section{Sgluon dipoles}
\label{sec:dipoles}

Sgluons are color octets with spin zero, so their dipoles are
identical to supersymmetric scalar quarks with modified color factors
$C_F \rightarrow C_A$. To remove the related infrared divergences we
implement the (un)integrated dipoles presented in Appendix~C of
Ref.~\cite{catani_seymour_massive} with this replacement. In addition,
we introduce a variable size of the subtraction phase space,
$0<\alpha\leq 1$, as pioneered in the FKS subtraction
scheme~\cite{alpha}. Values $\alpha < 1$ limit the phase space region
over which we subtract finite dipole contributions around the
soft-collinear pole.  Our notation closely follows
Ref.~\cite{catani_seymour_massive}.\medskip

From Eq.(C.1) of Ref.~\cite{catani_seymour_massive} we obtain the
sgluon dipole function $\left< V_{g\sg,k} \right>$ for the final-final
case.  The corresponding integrated dipole is decomposed into an
eikonal part including the soft integrals and the remaining hard
collinear integrals,
\begin{equation}
I_{g\sg,k} = 
C_A \left[ \, 2 I^\text{eik} + I_{g\sg,k}^\text{coll} \, \right] \; .
\label{eq:I_ijk}
\end{equation}
The divergent and finite parts of the regularized eikonal and
collinear integrals in $4-2\varepsilon$ dimensions are
\begin{alignat}{5}
\tilde{v}_{g\sg,k} \;
I^\text{eik}  = \; &
  \quad \frac{1}{2\varepsilon^2} \; 
         \left( 1-\left(\mu_\sg+\mu_k \right)^2 \right)^{-2\varepsilon}
         \left( 1-\frac{\rho_\sg^{-2\varepsilon}}{2} 
                -\frac{\rho_k^{-2\varepsilon}}{2} 
         \right) \notag \\
      &  +\frac{\zeta_2}{4}
          \left(6-\mu_\sg^{-2\varepsilon}-\mu_k^{-2\varepsilon}\right)
         +2 \text{Li}_2\left(-\rho\right)
         -2 \text{Li}_2\left(1-\rho\right) 
         - \frac{1}{2} \text{Li}_2\left(1-\rho_\sg^2\right)
         - \frac{1}{2} \text{Li}_2\left(1-\rho_k^2\right)
 \biggr] \notag \\
I_{g\sg,k}^\text{coll} = \; &
  \frac{2}{\varepsilon} 
 -\frac{\mu_\sg^{-2\varepsilon}}{\varepsilon} 
 -2\mu_\sg^{-2\varepsilon}
 +6
 -2\log \left(\left(1-\mu_k\right)^2-\mu_\sg^2\right)
 +\frac{4\mu_k\left(\mu_k-1\right)}{1-\mu_\sg^2-\mu_k^2} \; ,
\end{alignat}
where the rescaled mass $\mu_n$ and the variables $\rho$
and $\rho_n$, associated with the splitting $\tilde{ij}\rightarrow i\: j$
and the spectator $k$, are defined using the final state momenta
$p_i$, $p_j$ and $p_k$
\begin{alignat}{5}
\mu_n&=\frac{m_n}{\sqrt{Q^2}}
\qqqquad \text{with} \quad 
Q^\mu=p_i^\mu+p_j^\mu+p_k^\mu \notag \\
\rho&=\sqrt{\frac{1-\tilde{v}_{ij,k}}{1+\tilde{v}_{ij,k}}}
\qqquad \text{with} \quad 
\tilde{v}_{ij,k}=\frac{\sqrt{\lambda\left(1,\mu_{ij}^2,\mu_k^2\right)}}{1-\mu_{ij}^2-\mu_k^2}
\notag \\
\rho_n \left(\mu_j,\mu_k\right) &=\sqrt{\frac{1-\tilde{v}_{ij,k}+2\mu_n^2/\left(1-\mu_j^2-\mu_k^2\right)}
{1+\tilde{v}_{ij,k}+2\mu_n^2/\left(1-\mu_j^2-\mu_k^2\right)}} 
\qquad \left(n=j,k\right)  \;.
\end{alignat}
The kinematics of the splitting is described by 
\begin{equation}
\tilde{z}_j=1-\frac{p_ip_k}{p_ip_k+p_jp_k}
\qqquad 
y_{ij,k}=\frac{p_ip_j}{p_ip_j+p_ip_k+p_jp_k} \; ,
\end{equation}
where the upper limit in the $y_{ij,k}$ phase space integration is
\begin{equation}
y_{+}=1-\frac{2\mu_k\left(1-\mu_k\right)}{1-\mu_i^2-\mu_j^2-\mu_k^2} \; .
\end{equation}

To include the phase space parameter $\alpha$ in the
unintegrated dipole we simply add the step function $\Theta( \alpha -
y_{g\sg,k}/y_+)$.  It ensures that only for $y_{g\sg,k} < \alpha \;
y_+$ instead of the entire range $y_{g\sg,k} < y_+$ the dipole is
subtracted from the hard matrix element.  For the integrated dipole
part we modify Eq.\eqref{eq:I_ijk} by subtracting the finite phase
space contributions which a choice of $\alpha \ne 1$ removes,
\begin{alignat}{5}
I_{g\sg,k}\left(\alpha\right) & =  
 I_{g\sg,k}
-\triangle I_{g\sg,k}\left(\alpha\right) \notag \\
& =  I_{g\sg,k}
-\frac{2 \pi}{\alpha_{s}}
\int \left[dp_{g}\left(\tilde{p}_{g\sg},\tilde{p}_k\right)\right]
\Theta\left(\frac{y_{g\sg,k}}{y^+} - \alpha \right)
\frac{\left< V_{g\sg,k} \right>}{2p_{g}p_\sg} \; .
\label{eq:I_ijk_alpha}
\end{alignat}
For the (by definition) finite contribution $\triangle I_{g\sg,k}$ we
set $\varepsilon=0$.  The eikonal part of the kernel
$2/[1-\tilde{z}_\sg\left(1-y_{g\sg,k}\right)]$ is the same for the
$\left< V_{gQ,k} \right>$ and $\left< V_{g\sg,k} \right>$, so we use
the SM result for $\triangle I^\text{eik} (\alpha)$ as provided in
Eq.(A.9) of Ref.~\cite{bevilacqua}. The collinear part is different,
giving a correction to Eq.\eqref{eq:I_ijk} of the form
\begin{equation}
- \Delta I_{g\sg,k} \left(\alpha\right) = 
- C_A 
\left[ 2 \Delta I^\text{eik}\left(\alpha\right) 
+ \frac{1}{2\pi^2} 
\left( \frac{\left(1-\mu_k\right)^2-\mu_\sg^2}{1-\mu_\sg^2-\mu_k^2}
       \left(1-\alpha\right)
      +\log\alpha\right) 
\right] \; .
\end{equation}
As expected, this result becomes trivial for $\alpha =1$ and diverges in the limit
$\alpha \to 0$.\medskip

For the final-initial dipole function we start from Eq.(C.3) of
Ref.~\cite{catani_seymour_massive} which gives rise to the regularized
integrated dipole function
\begin{equation}
I_{g\sg}^a\left(x\right) =
C_A \left[ J_{g\sg}^a\left(x\right)_+
          +\delta \left(1-x\right) 
           \left( J_{g\sg}^{a;S} 
                 +J_{g\sg}^{a;NS}
           \right)
    \right] \; .
\label{eq:I_initial}
\end{equation}
The three contributions to $I_{g\sg}^a$ are
\begin{alignat}{5}
J_{g\sg}^a\left(x\right)_+ = & 
 \left( \frac{-2-2\log\left(1-x+\mu_\sg^2\right)}{1-x} \right)_+
+\left(\frac{2}{1-x}\right)_+ \log \left(2+\mu_\sg^2-x\right) \notag \\
J_{g\sg}^{a;S} = & 
 \frac{1}{\varepsilon^2}
-2 \zeta_2
-\mu_\sg^{-2\varepsilon}\left(\frac{1}{\varepsilon^2}
                          +\frac{1}{\varepsilon}
                          +\zeta_2
                          +2 \right)
-\frac{1}{\varepsilon} \log\left(1+\mu_\sg^2\right)
+\left(\frac{2}{\varepsilon}+4-\zeta_2\right) \notag \\
J_{g\sg}^{a;NS} = & 
 2 \zeta_2
-2\text{Li}_2\left(\frac{1}{1+\mu_\sg^2}\right)
-2\text{Li}_2\left(-\mu_\sg^2\right)
-\frac{1}{2}\log^2\left(1+\mu_\sg^2\right)
\end{alignat}
Again, we introduce an $\alpha$ parameter into the unintegrated
phase space integration, limiting the application of the dipole
subtraction to the region $1- x_{g\sg,a}<\alpha$. The kinematical 
variable $x_{ij,a}$ is given by
\begin{equation}
x_{ij,a}=\frac{p_ap_i+p_ap_j-p_ip_j+
\frac{1}{2}\left(m_{ij}^2-m_i^2-m_j^2\right)}{p_ap_i+p_ap_j}.
\end{equation}
The additional contribution to the integrated dipole 
$I_{g\sg}^a$ shown in Eq.\eqref{eq:I_initial} is
\begin{equation}
\triangle I_{g\sg}^a\left(\alpha\right) = 
C_A \frac{\Theta\left(1-\alpha-x \right)}{1-x}\left(-2+2\log\left(1+\frac{1}{1+\mu_\sg^2-x}\right)\right)
\end{equation}
Again, this result becomes trivial for $\alpha=1$ and
diverges in the limit of $\alpha\rightarrow0$ now when performing
the integral over $x$.\medskip

%------------------------------------------------
\begin{figure}[t]
\includegraphics[width=0.35\textwidth]{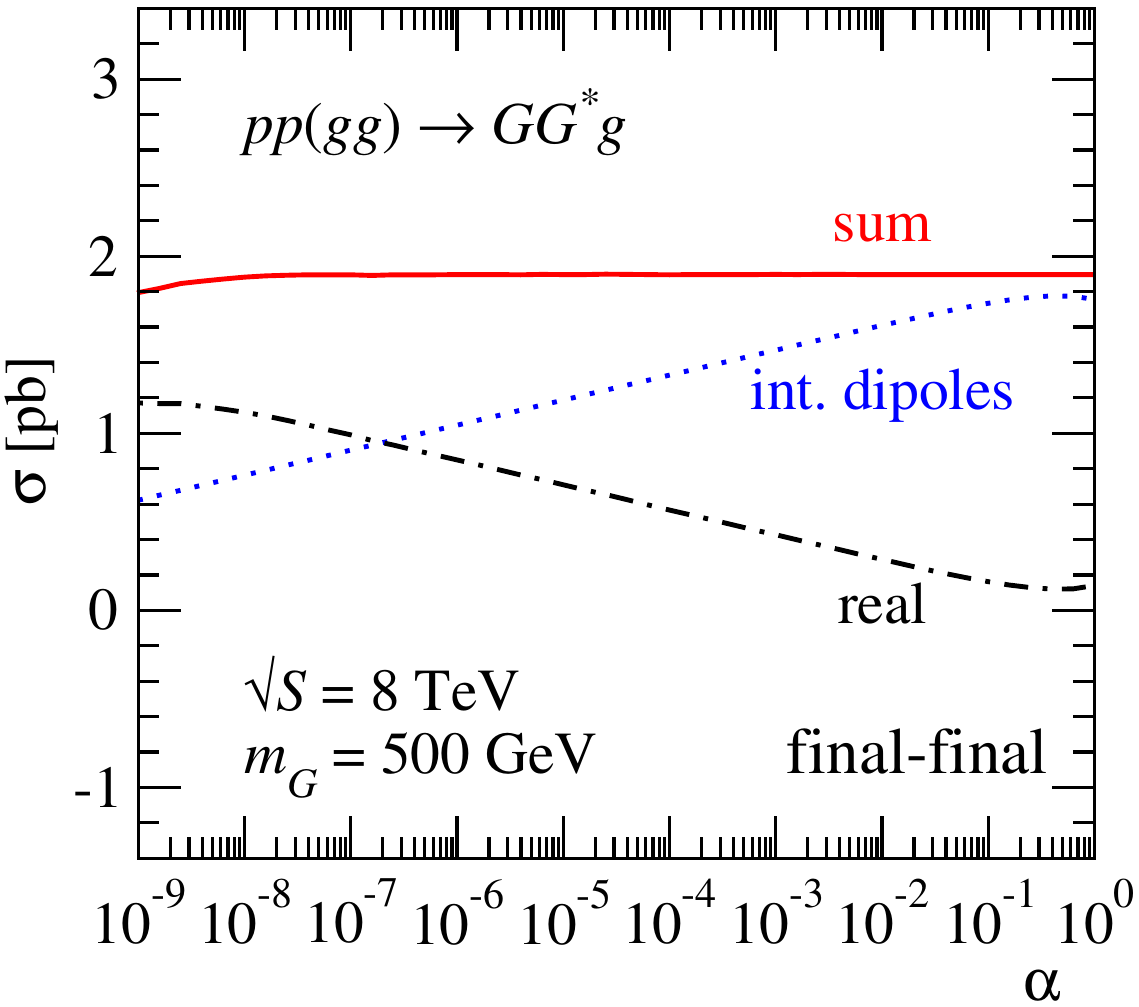}
\hspace*{0.1\textwidth}
\includegraphics[width=0.35\textwidth]{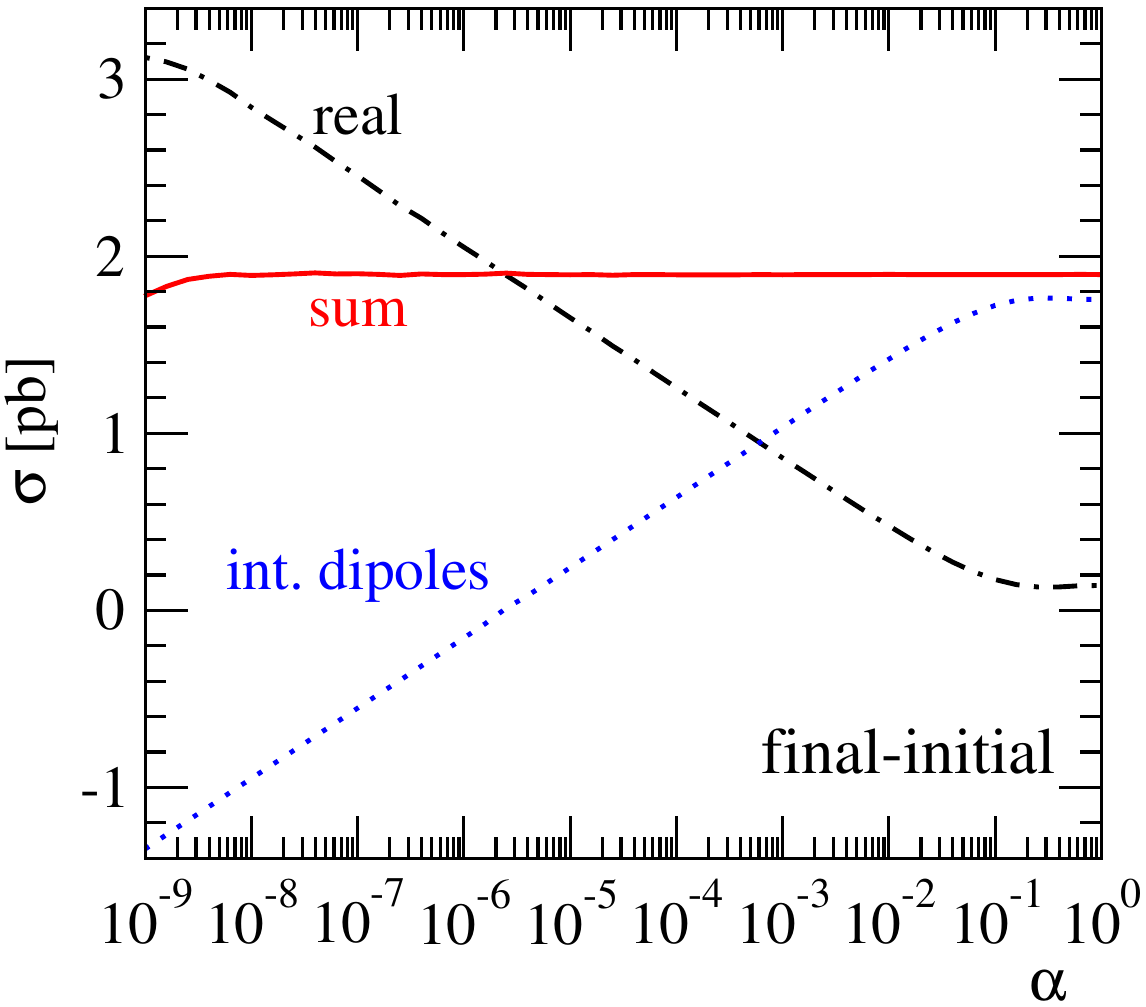}
\caption{$\alpha$ dependence of the final-final (left) and
  final-initial (right) sgluon dipoles for the sub-process $gg
  \rightarrow\sg\sg^*g$.}
\label{fig:alpha_IR}
\end{figure}
%------------------------------------------------

The numerical effects of a variable $\alpha$ parameter on the
subprocess $gg \rightarrow\sg\sg^*g$ we show in
Fig.~\ref{fig:alpha_IR}. While the individual unintegrated and
integrated dipole contributions diverge logarithmically with small
$\alpha$ the sum of them is numerically stable over eight orders of
magnitude. This kind of test is sensitive to many aspects of our
calculation, namely the proper coverage of all divergences, the
relative normalization of the two and three particle phase space,
etc. In Fig.~\ref{fig:alpha_IR} we see that a default value of $\alpha
= 10^{-3}$ gives roughly equal contributions from unintegrated and
integrated dipoles, avoiding large numerical cancellations for the
final-initial dipole. For the final-final dipole we would have to go
to smaller values of $\alpha$ which make the final-initial case
harder, so we use $\alpha = 10^{-3}$ throughout.

%%%%%%%%%%%%%%%%%%%%%%%%%%%%%%%%%%%%%%%%%%%%%%%%%%%%%%%%%%%%%%%%%%%%%%%%
\section{Renormalization} 
\label{sec:UV-renorm}

The ultraviolet counter terms we include automatically via the
leading-order {\sc Qgraf} output.  At present, {\sc MadGolem} fully
supports the calculation of NLO QCD corrections for the Standard
Model, the MSSM, and several other extensions of the Standard Model,
including sgluons.  For dimensional regularization we employ the
standard 't Hooft-Veltman scheme with $n=4-2\varepsilon$ dimensions.
The renormalization constants we define through the additive or
multiplicative relations between the bare and the renormalized fields,
\begin{equation}
\Psi^{(0)} \to Z^{1/2}_{\Psi}\,\Psi 
\qquad\qquad 
m_{\Psi}^{(0)}  \to m_{\Psi} + \delta m_{\Psi} 
\qquad\qquad 
g_s^{(0)} \to g_s + \delta g_s
\qquad (\text{with} \quad 
\Psi = q,A,\sg \,).
\label{eq:defRC}  
\end{equation}
These field, mass and coupling renormalization constants we
conventionally phrase in terms of two-point functions which we supply
in a separate library. Given a generic Lagrangian $\lag
(\Psi,m_{\Psi}, g_s)$ with a QCD interaction this consistently gives a
counter term Lagrangian of the form $\delta \lag (\Psi,m_{\Psi}, g_s,
\delta\Psi,\delta m_{\Psi}, \delta g_s)$.\medskip

\begin{figure}[t]
\includegraphics[width=0.3\textwidth]{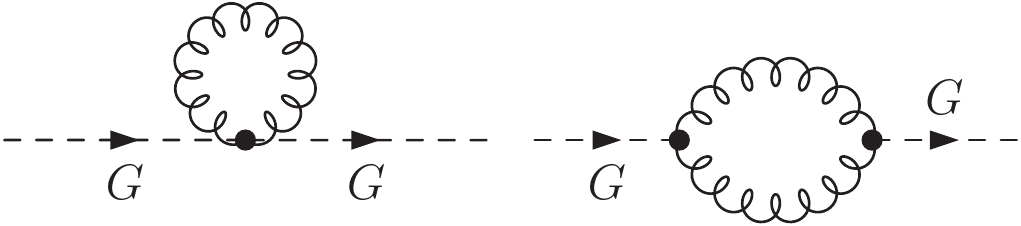} 
\hspace*{0.1\textwidth} 
\includegraphics[width=0.3\textwidth]{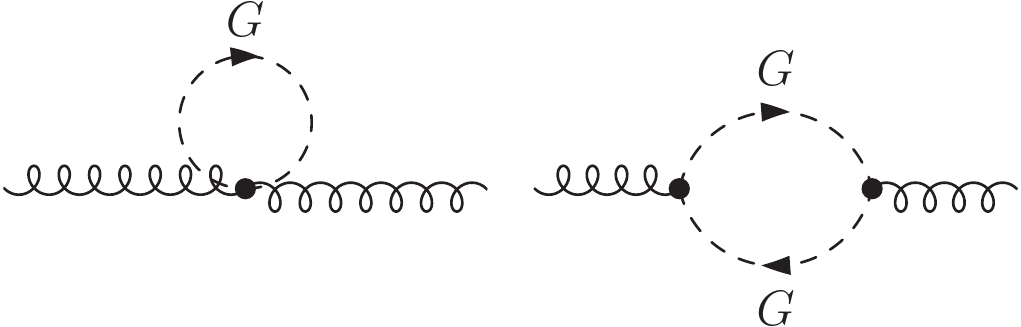} 
\caption{Feynman diagrams for the sgluon field renormalization (left)
  and sgluon-mediated gluon field renormalization (right).}
\label{fig:2point}
\end{figure}

First of all, a new sgluon field modifies the strong beta function.
If we start with the quantum corrections to the quark-quark-gluon
vertex in terms of the strong coupling $Z_{g_s}$, the gluon field
renormalization $Z_3$, and the quark field renormalization $Z_2$ this
translates into a combined $Z_1 = Z_{g_s}\,Z_2\,Z_3^{1/2}$.  Each of these
renormalization constants we expand as $Z_i = 1 + \delta_i +
\mathcal{O}(\alpha_s^2)$, with $\msbar$ counter terms $\delta_i$.  The
strong coupling constant renormalization at one loop we can thus write
as
\begin{alignat}{5}
\delta g_s = \delta_1 - \delta_2 - \frac{1}{2} \delta_3 & \notag \\
\text{with}\quad 
\delta_1 =& \delta_1^\text{SM} = -\frac{\alpha_s}{4\pi}\,(C_A+C_F)\,\Delta_{\varepsilon}
\notag \\
\delta_2 =& \delta_2^\text{SM} = -\frac{\alpha_s}{4\pi}\,C_F\,\Delta_\varepsilon
\notag \\
\delta_3 =& \delta_3^\text{SM} + \delta_3^\sg
         = \frac{\alpha_s}{4\pi}\, \left(\frac{5}{3}\,C_A - n_f\,C_F\,T_R\right)
          -\frac{\alpha_s}{12\,\pi}\,C_A\,\Delta_\varepsilon \; .
\label{eq:rengs1}
\end{alignat}
The shifted pole in the $\msbar$ prescription is $\Delta_{\varepsilon}
= 1/\varepsilon \, - \gamma_E + \log(4\pi)$ and the active
number of fermions is $n_f=6$. Because there are no direct couplings
between sgluons and matter fields $\delta_2$ keeps its SM value.  For
the same reason, sgluon-mediated corrections to the quark-quark-gluon
vertex are absent at one loop, so $\delta_1$ does not change. Only the
gluon self energy is modified by the triple and quartic gluon/sgluon
interactions, as displayed in Fig.~\ref{fig:2point}.

%------------------------------------------------
%------------------------------------------------

Combining all of the above contributions and decoupling the heavy
($H$) colored degrees of freedom --- in our case the top and the
sgluon --- gives us the final expression for $\delta g_s$ in terms of
the measured $\alpha_s$ values.  We implement this subtraction in the
zero-momentum scheme~\cite{decoupling,prospino_sqgl}. It
leaves the renormalization group running of $\alpha_s$ merely
determined by the light ($L$) degrees of freedom.  The renormalization
constant finally reads
\begin{alignat}{5}
\delta g_s &= 
- \frac{\alpha_s}{4\pi} \, \frac{\beta_0^L+\beta_0^H}{2} \, \Delta_\varepsilon 
- \frac{\alpha_s}{4\pi} \, 
  \left( \frac{1}{3}\,\log \frac{m_t^2}{\mu_R^2}
       + \frac{1}{2}\, \log \frac{m^2_\sg}{\mu_R^2} 
  \right) \notag \\ 
\beta_0 &= 
\beta_0^L + \beta_0^H =  
\left( \frac{11}{3}\,C_A - (n_f-1) \,C_F\, T_R \right) 
- \left( C_F\,T_R + \frac{1}{3}\,C_A \right) \; . 
\label{eq:alphas_ct}
\end{alignat}
\medskip
 
In a second step we need to compute the QCD renormalization constants
in the sgluon sector.  The sgluon two-point function receives
$\mathcal{O}(\alpha_s)$ corrections due to virtual gluon interchange,
as shown in Fig.~\ref{fig:2point}.  The corresponding ultraviolet divergences
we absorb into the sgluon mass $m_\sg$ and field-strength $Z_\sg$.  As
renormalization condition we choose the on-shell scheme
\begin{alignat}{7}
\Re\text{e}\; \hat\Sigma'(m_\sg^2) &= 0  
\qquad &\Rightarrow& \qquad 
\delta Z_\sg &=& -{\Re\text{e}\; \Sigma}'(m^2_\sg) \notag \\
\Re\text{e}\; \hat{\Sigma} (m_\sg^2) &= 0  
\qquad &\Rightarrow& \qquad 
\delta m_\sg &=&  + {\Re\text{e}\; \Sigma}(m^2_\sg) \;, 
 \label{eq:sgluonOS}
\end{alignat}
where $\Re\text{e}\,\hat{\Sigma}_\sg$ denotes the (real part of the)
renormalized sgluon self-energy,
\begin{equation}
 \hat{\Sigma}_\sg(q^2) = 
\Sigma_\sg(q^2) + (q^2 - m_\sg^2)\,\delta Z_\sg - \delta m_\sg^2 \; 
\label{eq:selfsgluon},
\end{equation}
and $\hat\Sigma'(q^2) \equiv d^2/dq^2\,\hat\Sigma(q^2)$ the
corresponding derivative with respect to the momentum squared.  The
analytic form of all renormalization constants we reduce down to one
and two-point scalar loop integrals~\cite{oneloop}. The sgluon mass
and field strength renormalization then reads
\begin{alignat}{5}
\delta Z_\sg &= 
\frac{\alpha_s}{2\pi}\, C_A\,
\left[B_0(m_\sg^2,m_\sg^2,0)+\,m^2_\sg\,B'_0(m_\sg^2, m_\sg^2,0) 
\right] \notag \\
\delta m_\sg &= 
-\frac{\alpha_s}{\pi}\,C_A\,
\left[\,m_\sg^2 + \frac{3}{4}\,A_0(m^2_\sg) 
\right] \; .
\label{eq:sgluonmass}
\end{alignat}
As expected, these expressions are identical to the squark case,
modulo a factor $C_A/C_F$ that reflects the different $SU_C(3)$
representations.

Finally, in Table~\ref{tab:cts} we quote the analytical expressions
for the relevant ultraviolet counter terms $\delta\lag$ as a function of the field,
mass, and strong coupling renormalization constants derived in this
Appendix.

%------------------------------------------------
\begin{table}[t]
\begin{center}
\begin{tabular}{ccl}  &  & \\ \hline & & \\
\myrbox{\includegraphics[scale=0.001]{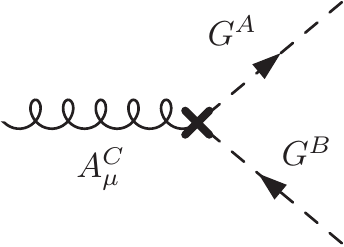}} & & 
$-i\,g_s\,f^{ABC}\,
 \left[\delta g_s + \frac{1}{2}\,
       \left(\delta Z_\sg + \delta Z_{\sg^*} + \delta Z_A\right)
 \right]\,\left[\sg^{*A}(\partial^\mu\,\sg^B)
- (\partial^\mu\,\sg^{*A})\sg^B\, \right]\,A^C_\mu$  \\ & & \\
 \\ & & \\ 
 \myrbox{\includegraphics[scale=0.001]{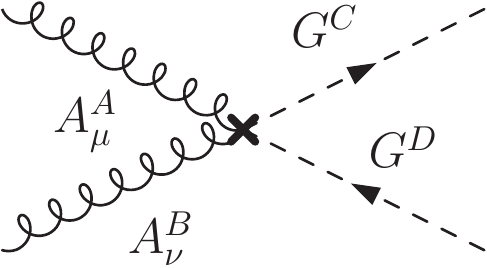}} &  & 
 $i\,g^2_s\,\left(f^{ACE}\,f^{BDE} + f^{ADE}\,f^{BCE}\right)\,
  \left[2\,\delta g_s + \delta Z_A +\delta Z_\sg \right]\,G^{*C}\,G^D\,A_\mu^A\,A^{B\,\mu}$
  \\ & & \\
   \myrbox{\includegraphics[scale=0.001]{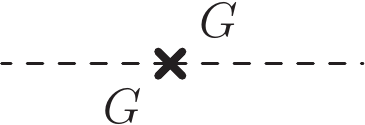}} &  & 
\quad $p^2\,\delta Z_\sg - \delta m^2_\sg\, - m^2_\sg\, \delta Z_\sg$
  \\ & & \\ \hline
\end{tabular}
\end{center}
\caption{Counter term Feynman rules for the sgluon-mediated interactions. \label{tab:cts}}
\end{table}
%------------------------------------------------

%%%%%%%%%%%%%%%%%%%%%%%%%%%%%%%%%%%%%%%%%%%%%%%%%%%%%%%%%%%%%%%%%%%%%%%%

\end{document}